\documentclass[10pt,twocolumn,letterpaper]{article}

\usepackage{iccv}
\usepackage{times}
\usepackage{epsfig}
\usepackage{graphicx}
\usepackage{amsmath}
\usepackage{amssymb}

\usepackage{booktabs}
\usepackage{multirow}

\usepackage[breaklinks=true,bookmarks=false]{hyperref}

\usepackage[accsupp]{axessibility}  

\iccvfinalcopy 

\def\iccvPaperID{****} 

\ificcvfinal\fi

\begin{document}

\title{\LaTeX\ Author Guidelines for ICCV Proceedings}

\title{T-AutoML: Automated Machine Learning for Lesion Segmentation using Transformers in 3D Medical Imaging}

\author{
    Dong Yang \quad
    Andriy Myronenko \quad
    Xiaosong Wang \quad
    Ziyue Xu \quad
    Holger R. Roth \quad
    Daguang Xu \\
    NVIDIA
}

\maketitle
\ificcvfinal\thispagestyle{empty}\fi

\begin{abstract}
Lesion segmentation in medical imaging has been an important topic in clinical research. Researchers have proposed various detection and segmentation algorithms to address this task.
Recently, deep learning-based approaches have significantly improved the performance over conventional methods.
However, most state-of-the-art deep learning methods require the manual design of multiple network components and training strategies.
In this paper, we propose a new automated machine learning algorithm,~\textbf{T-AutoML}, which not only searches for the best neural architecture, but also finds the best combination of hyper-parameters and data augmentation strategies simultaneously. The proposed method utilizes the modern transformer model, which is introduced to adapt to the dynamic length of the search space embedding and can significantly improve the ability of the search. We validate T-AutoML on several large-scale public lesion segmentation data-sets and achieve state-of-the-art performance.
\end{abstract}

\section{Introduction}
\begin{figure}[t]
    \centering
    \includegraphics[width=\linewidth]{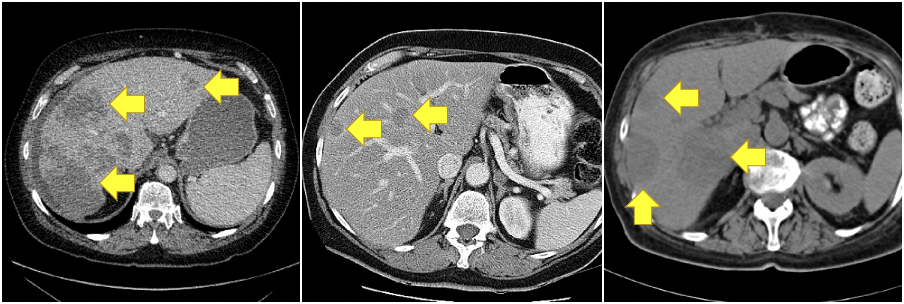}
    \caption{An example of intensity variations of liver lesions in 3D CT, which introduces substantial challenges for automated lesion segmentation. The yellow arrows indicate the target lesion locations. There are several large lesions in the left figure and small ones in the middle figure. The right figure presents an atypical CT contrast of liver and lesion regions after image windowing/normalization.}
    \label{fig:app}
\end{figure}

Nowadays, given the technological advances in algorithmic design (such as deep learning) and hardware platforms (such as GPU), medical image analysis has become a key step in disease understanding, clinical diagnosis, and treatment planning.
The sizes, shapes, and appearances of lesions in medical images can vary greatly across different anatomical structures (shown in Fig.~\ref{fig:app}). 
These semantic features are closely related to the severity of the disease.
At the same time, variations in the imaging patterns related to pathologies, scanning protocols, and medical devices introduce a computational challenge for automated algorithms~\cite{zhou2021review}. 
Therefore, automated lesion segmentation is one of the most challenging tasks in automated analysis of medical images, such as 3D CT, 3D MRI, histopathology, etc.
When facing with such challenges, most machine/deep learning models struggle to provide comprehensive solutions.

In recent years, deep convolutional neural networks have been widely applied for medical image analysis.
For instance, one of the most influential works in medical image segmentation using deep neural networks is U-Net~\cite{DBLP:conf/miccai/RonnebergerFB15,DBLP:conf/miccai/CicekALBR16}, which has been found to be more effective and efficient than previous non-learning based methods.
Furthermore, automated machine learning (AutoML) with neural networks has been explored for various applications, such as image recognition, semantic segmentation, object detection, natural image generation, etc.
One of the most popular tasks in AutoML is neural architecture search (NAS)~\cite{DBLP:conf/iclr/ZophL17}, which aims to design neural network architectures automatically without much human heuristics or assumptions.
Besides the model weights, the model structure itself becomes fitted for the task after searching, and is even transferable to different applications~\cite{DBLP:conf/cvpr/ZophVSL18}.
Additional constraints, such as latency or parameter quantity of models, can be added as searching objectives to fit the models into different computing platforms.
Several works~\cite{DBLP:conf/icml/TanL19} have achieved state-of-the-art  (SOTA) performance with NAS in large-scale image recognition (\eg, on ImageNet~\cite{DBLP:conf/cvpr/DengDSLL009}).
Other works applied similar techniques to optimize other components (\eg, augmentations, loss functions) or hyper-parameters in conventional deep learning solutions.
Although the recent development of AutoML has shown promising results in reducing the workload of algorithm design and improving model performance, most AutoML methods only optimize few components of the framework, \eg, network or transform, and continue to rely on the human choice of other components, which can lead to sub-optimal solutions.

In this paper, we propose a new method to automatically estimate ``almost'' all components of a deep learning solution for lesion segmentation in 3D medical images.
At first, a new search space for segmentation networks is introduced to enable flexible connection of global network structure beyond the U-shape design (encoder-decoder based models), which is commonly used in medical image segmentation~\cite{minaee2021image}.
Then, candidates of deep learning configurations (neural architecture, augmentations, hyper-parameters) are encoded to a 1D vector as the abstract representation.
Next, a binary relation predictor is trained with the representative vectors of configurations and their corresponding validating metrics.
More specifically, the predictor distinguishes between two input vectors to see if one could lead to better performance than the other.
Given such predictors, all configurations of deep learning solutions can be sorted with a direct comparison.
Finally, the search configurations are generated by sampling candidates from the candidate pool and picked by their predicted performance in the searched task of the lesion segmentation.
Furthermore, the searched configurations can be transferred to similar tasks in different datasets and achieve reasonable performance.

The contributions of our proposed method can be summarized as follows.
\begin{enumerate}
  \item We propose a new search space for lesion segmentation in 3D volumes, which has much more flexibility compared to the traditional U-shape architecture in medical imaging, and our architecture searching provides insights about the importance and fitness of various dense connections inside segmentation networks;
  \item Our method reaches the state-of-the-art performance of lesion segmentation on two public datasets;
  \item Both the search process and the configuration deployment of the proposed method is designed to be computationally efficient and effective;
  \item To the best of our knowledge, this is the first time that a comprehensive AutoML (including both NAS and optimization of other related training hyper-parameters) is applied to medical image segmentation leveraging the capacity of transformer modules.
\end{enumerate}


\section{Related Work}
\label{related_works}
\noindent\textbf{Lesion segmentation:}~Automated lesion segmentation in medical imaging has been studied for decades.
Some early studies involved handcrafted rules such as convergence index filter~\cite{MATSUMOTO2006343} and multi-scale hessian-based blob measurements~\cite{vos2012automatic}.
Later, researchers proposed to adopt machine learning models to tackle this challenging task.
The standard pipeline is to extract a series of handcrafted features, then distinguish lesion from the background by utilizing various classifiers, including the k-nearest neighbor~\cite{MURPHY2009757}, random forest~\cite{litjens2014computer,lay2018decomposable}, and support vector machine~\cite{kwak2015automated}.
The models can further leverage the advantage from ensembling with sufficient interpretation capacity. However, the reported model is usually very sensitive to imaging appearance with numerous false positives~\cite{lay2018decomposable}.
Recently, deep neural networks have been introduced to localize and segment lesions from multiple modalities, including MRI~\cite{schelb2019classification}, ultrasound~\cite{LIU2019101555}, and CT~\cite{WINKELS201915,DBLP:journals/mia/WangZLLGZDGT17}, given several publicly available annotated datasets.
The majority of these techniques take a network specifically designed for one previous task, and then do the training by pre-processing the data of the current task to accommodate the network design.

\noindent\textbf{Efficient ConvNet models for medical imaging:}~Researchers have adopted the design concepts of U-Net (``U''-shape convolutional encoder-decoder architecture with skip connections), and further extended it into novel and effective neural network architectures~\cite{DBLP:conf/3dim/MilletariNA16,DBLP:journals/tmi/LiCQDFH18,DBLP:journals/pami/BadrinarayananK17,badrinarayanan2017segnet,yang2017automatic,zhu2017unpaired,myronenko20183d,liu20183d,weng2019unet}.
Nowadays, 2D/3D U-Net is considered as the common baseline model for most segmentation tasks.
Often state-of-the-art performance in several tasks is reached via using variants of U-Net with model ensembling~\cite{DBLP:journals/corr/abs-1809-10486}.
However, one potential downside of U-Net is that once the U-shape models are trained properly, their prediction relies more on local context instead of the global structure because of the multi-level skip connections.
Then densely connected U-Net (\eg, U-Net$++$) was proposed to further improve the performance and generality of the segmentation models~\cite{DBLP:conf/miccai/ZhouSTL18}.
It gives us a hint that dense connection is helpful, but we have very limited intuition as to which connection plays a more important role and which connection can be pruned.
Recently, researchers used neural architecture search to answer such questions to improve the performance of 3D segmentation networks further.

\noindent\textbf{Automated machine/deep learning:}~Neural architecture search (NAS) is commonly used to design neural networks automatically with limited human heuristics to meet different user requirements (\eg, light-weight models or minimal computation)~\cite{elsken2018neural}.
Zoph and Le firstly introduced a novel framework to conduct neural architecture search using reinforcement learning for large-scale image recognition~\cite{DBLP:conf/iclr/ZophL17,DBLP:conf/cvpr/ZophVSL18,DBLP:conf/icml/PhamGZLD18}.
Specifically, they formulated the NAS problem as searching convolutional layers for repeated modules inside the entire network, which is capable of reducing the searching space and cost to a great extent since the same module structure is applied for several modules in the entire architecture.
Their searched networks (after being trained) are able to achieve state-of-the-art performance on the ImageNet dataset~\cite{deng2009imagenet}.
Others followed a similar idea to adopt reinforcement learning (RL) to tune the hyper-parameters of training settings or to find optimal data augmentation solutions~\cite{DBLP:conf/cvpr/CubukZMVL19,yang2019searching}.
Using newly searched neural networks, researchers have improved the performance of various computer vision applications (semantic segmentation, object detection, design of loss function, etc.) in some challenging datasets~\cite{bello2017neural,chen2018searching,xu2018autoloss,dong2019neural,ghiasi2019fpn}.
Meanwhile, researchers have explored other concepts of NAS, \ie, ``super-net'' and one-shot NAS~\cite{cai2018proxylessnas,DBLP:conf/iclr/LiuSY19,xia20183d,liu2019auto,guo2019single,wu2019fbnet,shaw2019squeezenas,xie2019exploring,gong2019autogan}.
Another trend in NAS is predictor-based methods~\cite{DBLP:conf/cvpr/WangWCLL0LH20,DBLP:conf/nips/DudziakCALKL20,DBLP:conf/eccv/YuJLBKTHSPL20}.
The predictors do not have performance gaps of searched architectures between searching and re-training.
However, they require many trained models with validation accuracy as the ground truth to train a robust predictor model.

Recently, the related research has been widely developed in automated deep learning (AutoDL) to optimize almost every aspect of deep learning, including network architecture design~\cite{DBLP:conf/iclr/ZophL17,real2019regularized,liu2019auto,DBLP:conf/icml/TanL19,DBLP:conf/icml/WistubaP20,DBLP:conf/icml/WistubaP20}, data augmentation strategies~\cite{DBLP:conf/cvpr/CubukZMVL19,yang2019searching,DBLP:conf/cvpr/MittalLKFCF20}, and loss functions~\cite{DBLP:journals/corr/abs-1810-02442}, etc.
With those automated components of deep learning, several computer vision applications successfully perform much better in various metrics (accuracy, latency, model size, etc.).
However, previous works mainly focus on optimizing few components of a deep learning framework only.
Our proposed framework aims to simultaneously search for the optimal combination of deep learning components, introducing greater flexibility to AutoDL.

\section{T-AutoML}
\label{method}
\begin{figure*}[h]
    \centering
    \includegraphics[width=0.9\linewidth]{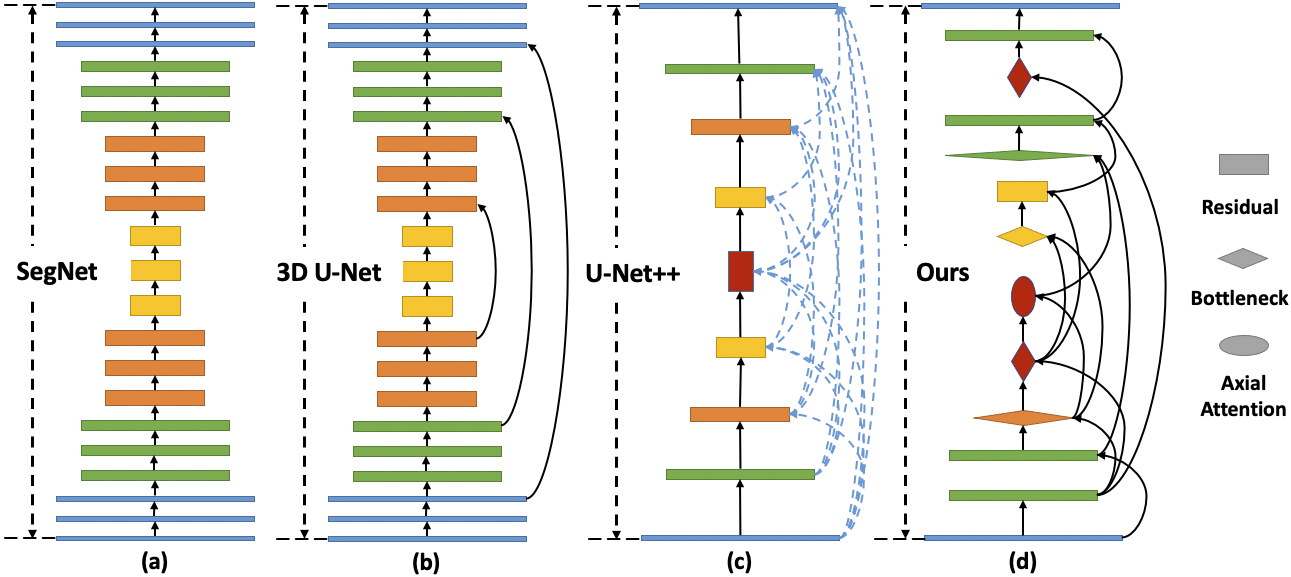}
    \caption{(a) SegNet~\cite{badrinarayanan2017segnet}; (b) U-Net~\cite{DBLP:conf/miccai/CicekALBR16} (c) U-Net$++$~\cite{DBLP:conf/miccai/ZhouSTL18}; The connected dotted line shows connectivity, but the underlying operation is more complicated than a single convolution. (d) Our searched the architecture. Different colors indicate different spatial resolutions. Different block types are marked as different block shapes.}
    \label{fig:net}
\end{figure*}

The success of deep learning comes from several aspects.
One major advantage of deep learning is that neural networks are end-to-end trainable without the need for feature engineering.
However, there is a new need for designing the best network architecture, which often determines the upper-bound performance of deep learning models.
A good architecture enables effective gradient back-propagation during training and feature learning.
To further promote the performance, data augmentation during training is utilized to increase model robustness and mitigate the gap between the domains of training, validation, and testing datasets.
Last but not least, hyper-parameters in model training are also critical for fast convergence and decent accuracy.

Our proposed transformer-based AutoML (T-AutoML) method (see Sec.~\ref{sec:neural_predictor}) trains a predictor to compare performance between different training configurations (neural architecture, data augmentation, and hyper-parameters).
T-AutoML leveraging the advanced capacities of the transformer modules~\cite{DBLP:conf/nips/VaswaniSPUJGKP17} to handle the input sequence with dynamic sequence, which is suitable for neural architecture space with a various number of blocks/layers.
The proposed method is capable of covering (almost) all components in a conventional deep learning solution.
The neural network architecture, data augmentation, and other related hyper-parameters can be fit into the method with proper encoding strategies.
And the combination of encoding is the reference for the predictor to determine the optimal architecture and training configurations for the target tasks.
\subsection{SpineNet Searching Space}
U-shape neural networks for segmentation are very popular in modern deep learning since they are good at image-to-image translation~\cite{DBLP:journals/pami/BadrinarayananK17,DBLP:conf/miccai/RonnebergerFB15,DBLP:conf/miccai/CicekALBR16,DBLP:conf/3dim/MilletariNA16,DBLP:journals/corr/abs-1809-10486}.
The encoder reduces the feature maps gradually, and meanwhile increases the filter numbers.
The decoder reverts back the spatial resolution gradually to match the input shape.
Also, the skip connections between the encoder and decoder in the U-shape have been validated to be effective in capturing the low-level image details for the fine segmentation boundaries~\cite{drozdzal2016importance}.
Although such a design is simple and straightforward, it produces promising results in many segmentation applications, and has become the go-to network in the field of medical image segmentation.
However, there is no evidence that if the ``encoder, decoder, and skip connection'' design patterns are optimal for the task at hand.

Hence, we propose a new search space of neural architectures to further study the network connections for image segmentation.
Our method is inspired by SpineNet~\cite{DBLP:conf/cvpr/DuLJGTCLS20}, which was proposed for object detection in natural image processing.
Under the framework, the feature maps at different spatial levels of the network can be connected with each other arbitrarily.
At the same time, the order of operations to increase or decrease the feature maps' spatial size can be arranged randomly.
Thus, the search space not only contains U-shape networks or densely connected networks~\cite{DBLP:conf/miccai/ZhouSTL18} but also covers other network topologies with asymmetric structure (see Fig.~\ref{fig:net}).
Therefore, the search space is much larger than previous ones in the segmentation-related NAS literature~\cite{DBLP:conf/cvpr/LiuCSAHY019,DBLP:conf/miccai/BaeLLPCJ19,DBLP:conf/miccai/KimKLBKCYK19,DBLP:conf/3dim/ZhuLYYX19,DBLP:conf/cvpr/YuYRBZYX20,DBLP:conf/miccai/JiZLRZL20}.

Like other methods in NAS, the search space consists of several blocks with different operations.
In our method, the candidates of different blocks are 3D residual blocks, 3D bottleneck blocks, and 3D axial-attention blocks~\cite{DBLP:conf/eccv/WangZGAYC20}.
The residual and bottleneck blocks are effective in avoiding vanishing gradient~\cite{DBLP:conf/cvpr/HeZRS16}.
The axial-attention block was originally proposed to resolve the issue of weak long-range dependency in the 2D plane for segmentation tasks~\cite{DBLP:conf/eccv/WangZGAYC20}.
Our axial-attention is the extension from 2D to two 3D versions.
The first one is conducted axis-wise attention along $\mathcal{X}$, $\mathcal{Y}$, $\mathcal{Z}$ axes sequentially.
And the second one is conducted along $\mathcal{Y}$, $\mathcal{X}$, $\mathcal{Z}$ axes sequentially.
Since the axial plane ($\mathcal{X}$-$\mathcal{Y}$ plane) is commonly imaged in higher resolution and is more informative with low-level image details compared to $\mathcal{X}$-$\mathcal{Z}$ plane and $\mathcal{Y}$-$\mathcal{Z}$ planes, we let the axial attention block process the $\mathcal{X}$-$\mathcal{Y}$ plane first, and then conduct $\mathcal{Z}$ axis-wise attention.

In practice, we construct the architecture with $\mathcal{N}$ blocks one-by-one.
Whenever introducing a new block $c$ to the architecture, its category and spatial resolution level of the block needs to be determined first.
From the 3rd block, the block $c_i$ would collect feature maps from two of all precedent blocks $c_j$ and $c_k$, and combine them to a single feature map
($c_2$ would only receive the feature map from $c_1$. $i$, $j$, $k$ are not necessarily adjacent).
In order to combine layers from different spatial resolutions and match the resolution of the current block, necessary up-sampling and down-sampling are applied to the feature maps, respectively.
We change the spatial resolution of combined feature maps to the target spatial resolution following the fashion from~\cite{DBLP:conf/cvpr/DuLJGTCLS20}.
Then the combined feature maps would be converted to $c_i$'s spatial resolution level with necessary up-sampling/down-sampling.
Lastly, the $\mathcal{N}$-th block is the one before the final activation layer (softmax layer to generate multi-class probability maps).
During the search process, $\mathcal{N}$ and spatial resolution can be determined within certain ranges of discrete integer values.
The connection between previous blocks and the current one can be arbitrary.
In order to further reduce GPU memory consumption during training and meanwhile speed-up the training process, we have one stem layer to down-sample input volume by half of its original size using $3 \times 3 \times 3$ convolution following~\cite{liu2019auto}.
At the end of the architecture, another up-sampling layer (linear interpolation) is used to restore the feature maps back to the original volume size.
\subsection{Encoding}
To represent architecture and other training configurations and simplify the computation of the next step, we encode architecture and training configuration in the search space jointly to form a ``large'' one-dimensional vector $\mathcal{V}$.
$\mathcal{V}$ encodes both numerical values and non-numerical values (\eg, choices of optimizers/losses/data augmentation).

The architecture is encoded as a one-dimensional vector $\mathcal{A}$ with dynamic length.
For each block, we use five integer indices to represent its current block ID, choice of operations, spatial resolution level, and two IDs of predecessor blocks, respectively.
The IDs of predecessor for the 1st block is $\left ( -1,-1 \right )$, for the 2nd block is $\left ( 0,-1 \right )$.

During training, we apply $n=5$ augmentation methods in a sequence.
Therefore, we have $n$ place holders for $m$ augmentation candidates.
For each place holder, we use indices (0 to $m-1$) to indicate the choice of augmentation method.
Thus, the 1D vector for augmentation has length $n$.
Meanwhile, we encode the options of different optimizers and loss functions using integer indices as well.
The other related hyper-parameters (\eg, learning rate) can be further optimized as long as they can be formulated into continuous or discrete values.
After encoding all necessary components in the search space, we concatenate all 1D vectors into one large vector $v$.

\noindent\textbf{Search space:} The search space is designed to cover most components in a typical deep learning framework.
For data augmentation, the candidates are random flipping (along $\mathcal{X}$, $\mathcal{Y}$, $\mathcal{Z}$ axes respectively), random rotation (90 degrees) in $\mathcal{X}$-$\mathcal{Y}$ planes, random zooming, random Gaussian noise, random intensity shift, random intensity scale shift.
By default, we set the probability of activation (of each augmentation) to 0.15 according to the recommendations in~\cite{isensee2021nnu}.
The candidate learning rates are $\left[0.01,0.005,0.001,0.0005,0.0001\right]$, and the candidate learning rate schedulers include constant and polynomial scheduler.
The loss function candidates can be determined from (soft) Dice loss~\cite{DBLP:conf/3dim/MilletariNA16} with or without squared prediction, cross entropy (CE) loss, combinations of Dice loss and CE loss, and combinations of Dice loss and focal loss~\cite{DBLP:conf/iccv/LinGGHD17,zhu2018anatomynet}.
The optimizer candidates are Adam, stochastic gradient descent, momentum, Nesterov and NovoGrad~\cite{DBLP:journals/corr/abs-1905-11286} optimizers.
For the architecture space, the number of blocks $\mathcal{N}$ can be selected from 5 to 12 and the spatial resolution level $l$ can be 2 to 5.
At each spatial level, the spatial size of the feature maps is $1/2^{\left ( l-1 \right )}$ and the number of channels is $2^{\left ( l-1 \right )\cdot c_1}$.
Here, $c_1$ is set to be 16.
The block candidates are 3D residual block, 3D bottleneck block, and 3D axial-attention blocks.
\subsection{Neural Predictor for Binary Relationship}
\label{sec:neural_predictor}
\begin{figure}[t]
    \centering
    \includegraphics[width=0.8\linewidth]{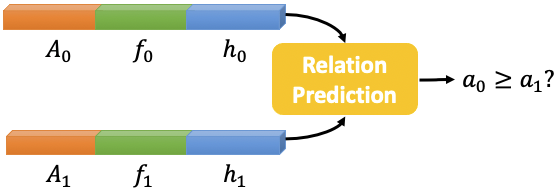}
    \caption{The predictor takes input vectors $v_0$ and $v_1$ (containing encodings of architecture $A$, augmentation $f$, and hyper-parameters $h$), and predicts which vector would generate higher validation scores $a$.}
    \label{fig:pred}
\end{figure}

The performance predictor takes the 1D encoding vector $\mathcal{V}$ containing neural architecture, data augmentation, and hyper-parameters, and outputs the corresponding performance (validation accuracy, etc.).
Such a predictor is able to cover all possible components in machine/deep learning, and it can be potentially transferred across different datasets, tasks, and hardware platforms.
However, the main drawback of the predictor-type AutoML methods is that they require a lot of training time to generate ground truth in order to train a stable predictor.

To alleviate the burden of training many jobs/instances, we proposed a new predictor-based search method to predict the relation between different configuration vectors $v_i$ and $v_j$ instead.
The goal is to predict the relation of validation accuracy $a$ between two configurations $g_i$ and $g_j$: better or worse (\ie, larger or smaller accuracy) shown in Fig.~\ref{fig:pred}.
After vector $v$ is extracted from the raw configuration, we can adopt transformer modules~\cite{DBLP:conf/nips/VaswaniSPUJGKP17} and fully-connected (FC) layers mapping the vector to a binary prediction.
The transformer encoder is adopted to encode the entire vector $v$ with dynamic length into feature maps with a fixed size.
Multiple FC layers convert the high-level feature maps into binary relation predictions.
The ground truth $\mathrm{GT}_{i,j}$ for training such a predictor is based on better or worse validation scores $a_i$, $a_j$ as in Eq.~\ref{eq:gt}.
\begin{equation}
    \mathrm{GT}_{v_i,v_j}=\left \{ \begin{matrix}
 1& a_i \geq a_j\\ 
 0& a_i < a_j
\end{matrix} \right.
\label{eq:gt}
\end{equation}
We formulate the predictor as a binary classifier instead of an accuracy regressor.
Once the predictor is trained, it can be used to rank unseen configurations with a sorting algorithm.
For each configuration, we compare it with all other sampled configurations.
It would be indexed with how many configurations have worse validation accuracy than it.
And sorting can be simply conducted based on such indices.
The comparison between $v_i$ and $v_j$ is light-weighted using CPU or GPU.
Thus, the sorting would be finished in seconds for hundreds of randomly sampled candidates.

The advantage of this predictor design is that it requires less overall training time.
Firstly, various configurations can be compared with fewer training iterations, since the absolute values of predicted accuracy are not always necessary.
Especially when transferring the predictor to another dataset, the predicted accuracy becomes much less informative.
The actual ranking between configuration $v_i$ and $v_j$ is more informative to a new task without any search/training experience, compared to the predicted accuracy with the previous predictor.
Second, achieving a stable predictor (accuracy regressor) requires at least hundreds of ground truth (GT) data points, which also increases the overall time cost~\cite{DBLP:conf/cvpr/WangWCLL0LH20}.
In our method, the binary relation predictor requires much fewer GT points in order to learn a similar amount of parameters of the predictors.
For instance, training 20 jobs can create only 20 GT points for an accuracy-based predictor.
However, the same amount of trained jobs/instances can create $20 \times 20 = 400$ GT points for the relation-based predictor.
Therefore, such a relation can be estimated with less training time/iterations.

\section{Experiments}
\label{experiments}

\subsection{Datasets}

\noindent\textbf{LiTS 2017:} The LiTS challenge is about developing automatic segmentation algorithms to segment liver lesions in contrast-enhanced abdominal CT scans~\cite{DBLP:journals/corr/abs-1901-04056}.
It has been publicly available since 2017, and more than 3000 participants in the challenges made their submissions over the years.
CT volumes and ground truth labels are provided by multiple clinical sites around the world.
The dataset contains 201 3D abdominal CT volumes with ground truth labels of liver and lesion segmentation.
131 of them are used for model training, and 70 volumes are used for testing on a public server.
The ground truth labels of the test set are not visible to the participants.
The spacing of 3D CT volumes varies from $0.598 \times 0.598 \times 0.450$ $mm^3$ to $0.977 \times 0.977 \times 6.0$ $mm^3$, and the range of volume shapes is from $512 \times 512 \times 42$ $\mathrm{voxel}^3$ to $512 \times 512 \times 1026$ $\mathrm{voxel}^3$.
Large variance also exists in the imaging quality, contrast, and field-of-view (FOV) (shown in Fig.~\ref{fig:app}).
The dataset has been largely adopted for performance evaluation of 3D segmentation methods by many established works.
In practice, the image intensity is clipped within an abdominal window of $\left[-125,225\right]$ Hounsfield units (HU) and pre-processed using standard normalization.
The dataset is re-sampled into the voxel spacing $\left ( \min\left ( 1.0,s_x \right )~mm, \min\left ( 1.0,s_y \right )~mm, 1.0~mm\right )$ based on each case's spacing $s$.
Such re-sampling strategies are proved to be capable of preserving some small lesions in the high-resolution CT images.

\noindent\textbf{Medical Segmentation Decathlon:}~The medical decathlon challenge (MSD) hosts several tasks of 3D medical image segmentation~\cite{msd2018}.
We choose to take 06 (lung lesion/tumor segmentation in 3D CT images) to test the transferability of our searched configurations from previous LiTS datasets.
The dataset contains 63 volumes for training and validation, and 32 volumes for testing.
The spacing of 3D CT volumes varies from $0.598 \times 0.598 \times 0.625$ $mm^3$ to $0.977 \times 0.977 \times 2.5$ $mm^3$, and the range of volume shapes is from $512 \times 512 \times 84$ to $512 \times 512 \times 947$.
The image intensity is clipped in a chest window of $\left[-1000,1000\right]$ HU and pre-processed using linear mapping to $\left[0,1\right]$.
The dataset is re-sampled into the isotropic voxel spacing 1.0 $mm$.

\noindent\textbf{Implementation:}~The task for searching is the liver and lesion segmentation on LiTS dataset.
The segmentation model takes a 1-channel input and outputs 3-class probability maps (background, liver, and lesion, respectively) with the same shape as the input.
The task 06 of MSD is utilized to verify the transferability of our searched deep learning configurations from LiTS.
The task is formulated as a binary segmentation problem.
The last layer of search network is modified to take a 1-channel input and output 2-class probability maps (background, lung lesion).

To train and validate the predictor, we uniformly sample 100 configuration candidates from the search space.
75 of them are used for predictor training, and the remaining 25 are used for validation.
Then, we train each candidate configuration for 10,000 iterations, and validate the segmentation model every 1,000 iterations.
The best validation Dice score is referred to as the GT accuracy for that configuration.
Once all the GT points are generated, the predictor model would be trained for 10,000 iterations with Adam optimizer, learning rate 0.001, and batch size 32.
At last, we select the optimal configuration from the predictor with 200 candidates (100 existing ones and another 100 unseen random samples) to be our final model training solution.
We use single GPU training jobs/instances for the configuration search.
Each job takes about 3 hours for both training and validation, and the total searching time is around 300 GPU hours for the task.
But our search process can be fully parallel, and thus searching would be done within two days using an 8-GPU server.
Our searching efficiency is decent considering the huge search space.
For example, C2FNAS used thousands of GPU hours to search for neural architectures (3D segmentation network) only~\cite{DBLP:conf/cvpr/YuYRBZYX20}.
The predictor model takes few minutes to train.

Our searched neural network is shown in Fig.~\ref{fig:net}.
It has the least amount of parameters (16.96M), compared with other popular network models (19.07M for 3D U-Net~\cite{DBLP:conf/miccai/CicekALBR16}, 22.60M for 3D U-Net$++$~\cite{DBLP:conf/miccai/ZhouSTL18}, 17.02M for C2FNAS~\cite{DBLP:conf/cvpr/YuYRBZYX20}).
We reuse the configuration for Tab.~\ref{tab:config} from the search method to determine all necessary components.
During training, the input of the network are patches with size $128 \times 128 \times 128$, and the ratio between foreground and background patches is $1:1$.
The batch size is 4 (2 patches from 2 volumes) per GPU.
To achieve better and robust segmentation performance, we linearly extend the number of total training iterations to 40,000 with the same learning rate scheduler as the searched one.
The validation is conducted per 1,000 iterations to select the best model checkpoint.
The validation accuracy is measured with the Dice score.
The model inference uses a sliding-window scheme, and the overlapped region of adjacent windows is 80\% of the window size.
Similar to~\cite{DBLP:journals/corr/abs-1809-10486,isensee2021nnu}, we conducted 5-fold cross-validation for all tasks, resulting in 5 segmentation models after training.
The final prediction of test data is the ensemble result of the probability maps from the 5 models.
Our proposed method is implemented with PyTorch and trained on two NVIDIA V100 GPUs with 16 GB memory.

\begin{table}[h]
\begin{center}
\begin{tabular}{lc}\toprule
\multicolumn{1}{l}{\textbf{component}} & \textbf{searched result} \\\midrule
data augmentation             & \begin{tabular}[c]{@{}c@{}}random flip, \\ random gaussian noise,\\ random intensity shift,\\ random zoom,\\ random intensity scale shift\end{tabular} \\\midrule
learning rate (LR)            & 0.0001                                                                                                                                                 \\\midrule
LR scheduler                  & constant                                                                                                                                               \\\midrule
loss function                 & \begin{tabular}[c]{@{}c@{}}Dice loss (no squared prediction)\\ + CE\end{tabular}                                                                       \\\midrule
optimzer                      & Adam \\
\bottomrule
\end{tabular}
\end{center}
\caption{The final searched training configuration for liver and lesion segmentation using LiTS dataset.}
\label{tab:config}
\end{table}
\subsection{Results}
\begin{figure*}[h]
    \centering
    \includegraphics[width=\linewidth]{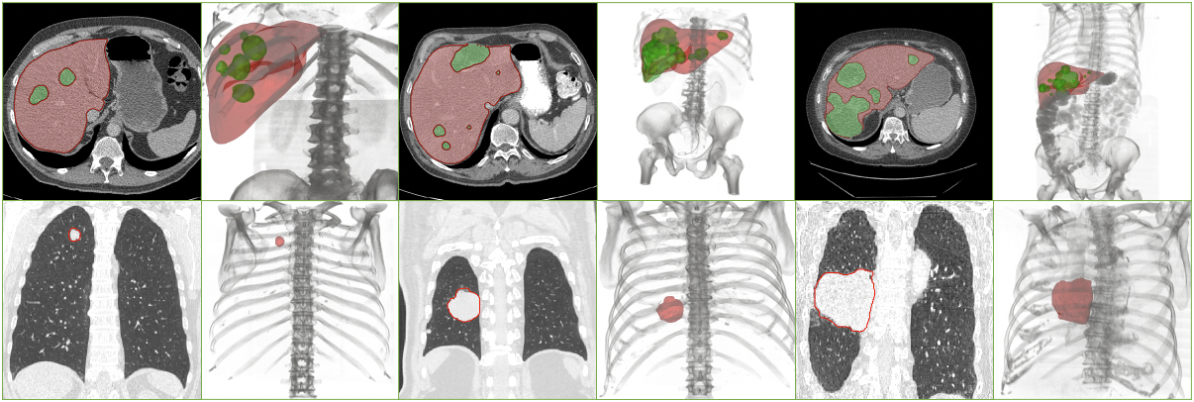}
    \caption{The first row shows the superimposed display of the liver and lesion segmentation in CT and their 3D rendering results. The second row shows the superimposed display of lung lesion segmentation in chest CT and its 3D rendering.}
    \label{fig:vis}
\end{figure*}

\noindent\textbf{Evaluation:}~We evaluate the segmentation accuracy in terms of the Dice score per case for LiTS challenge, and the Dice score and the normalized surface distance (NSD) for the lung task of the MSD challenge. 
Tab.~\ref{tab:lits}, Tab.~\ref{tab:task06} and Fig.~\ref{fig:vis} results demonstrate the superior performance of our method compared to the other SOTA methods. 

For the LiTS dataset, our method performs better for both lesion and liver segmentation. Tab.~\ref{tab:lits} shows that our method was able to capture not only the lesion regions, but also the fine boundary of the large organs.
Unlike other works, our method does not use any advanced model training features, such as sophisticated boundary loss functions~\cite{zhang20203d} or deep supervision~\cite{DBLP:journals/corr/abs-1809-10486}.
Our method benefits from fundamental components of neural network models and the training process.
For the lung task of the MSD challenge, our method performs better than other SOTA methods with substantial improvements in terms of both the Dice score and the NSD.
In the evaluation system of MSD, the higher the NSD value is, the better the segmentation quality is provided by the models.
By inspecting the MSD leaderboard statistics of other methods for failed cases, we observe that all previous methods tend to make false negative predictions in a few specific test volumes (because the min metric values are close to 0).
Whereas our method succeeds in such cases by capturing lesion regions well.
Moreover, our method has clear advantage in 25 percentile and standard deviation of overall metric values, which shows our proposed method has both accurate and robust performance in this task.
The fact that we migrated the found solution from LiTS to MSD lung task without any additional search, validates the transferability of our method.


\begin{table}[h]
\begin{center}
\begin{tabular}{lccc}\toprule
       & lesion & liver & mean \\ \midrule
RA-UNet~\cite{DBLP:journals/corr/abs-1811-01328} & 0.5950 & 0.9610 & 0.7780 \\
AH-Net~\cite{DBLP:conf/miccai/LiuXZPGMWJCC18} & 0.6340 & 0.9630 & 0.7895 \\
FCN~\cite{DBLP:conf/isbi/VorontsovTPK18} & 0.6610 & 0.9510 & 0.8060 \\
3D-DenseUNet~\cite{alalwan2021efficient} & 0.6960 & 0.9620 & 0.8290 \\
H-DenseUNet~\cite{DBLP:journals/tmi/LiCQDFH18} & 0.7220 & 0.9610 & 0.8415 \\
LW-HCN~\cite{DBLP:conf/ijcai/ZhangXZCXS19} & 0.7300 & 0.9650 & 0.8475 \\
$\mathrm{U}^3$-Net~\cite{DBLP:journals/access/TranCL21} & 0.7369 & 0.9638 & 0.8504 \\
Cascade U-ResNets~\cite{DBLP:journals/access/XiWSCFH20} & 0.7520 & 0.9490 & 0.8505 \\
VolumetricAttention~\cite{DBLP:conf/miccai/0007HCGV19} & 0.7410 & 0.9610 & 0.8510\\
MA-Net~\cite{DBLP:journals/access/FanWLW20} & 0.7490 & 0.9600 & 0.8545 \\
DistanceMetric~\cite{zhang20203d} & 0.7640 & 0.9650 & 0.8645 \\
nnU-Net~\cite{isensee2021nnu} & 0.7630 & 0.9670 & 0.8650 \\
\midrule
 T-AutoML (ours)   & \textbf{0.7650} & \textbf{0.9670} & \textbf{0.8660} \\
\bottomrule
\end{tabular}
\end{center}
     \caption{LiTS challenge test-set performance evaluation for lesion and liver segmentations in terms of the average Dice score per case. 
     The metrics of our method are copied from the LiTS  leaderboard, and the metrics of the other methods are copied from their respective publications and the leaderboard entries.
     }
     \label{tab:lits}
\end{table}


\begin{table}[h]
\begin{center}
\begin{tabular}{lcccc}
\toprule
                     & \multicolumn{4}{c}{Dice}                                                                  \\\cline{2-5} 
                     & mean                 & std                 & min.               & 25\%                 \\
\midrule
MPUnet~\cite{DBLP:conf/miccai/PerslevDPI19} & 0.5900 & - & - & -\\
C2FNAS~\cite{DBLP:conf/cvpr/YuYRBZYX20} & 0.7044 & 0.2099 & 0.0022 & 0.6042\\
nnU-Net~\cite{isensee2021nnu} & 0.7397 & 0.2164 & 0.0000 & 0.6041 \\
\midrule
 T-AutoML (ours) & \textbf{0.7533} & \textbf{0.1574} & \textbf{0.3530} & \textbf{0.7014}\\
\midrule
                     & \multicolumn{4}{c}{NSD}                                                                   \\\cline{2-5} 
                     & mean                 & std                 & min.                & 25\%                 \\
\midrule
MPUnet~\cite{DBLP:conf/miccai/PerslevDPI19} & 0.5600 & - & - & -\\
C2FNAS~\cite{DBLP:conf/cvpr/YuYRBZYX20} & 0.7222 & 0.2897 & 0.0031 & 0.5116 \\
nnU-Net~\cite{isensee2021nnu} & 0.7602 & 0.2962 & 0.0000 & 0.7018\\
*U-Net$++$~\cite{DBLP:conf/miccai/ZhouSTL18} & 0.7721 & - & - & -\\
\midrule
 T-AutoML (ours) & \textbf{0.7768} & \textbf{0.2816} & \textbf{0.0998} & \textbf{0.7392} \\
\bottomrule
\multicolumn{1}{l}{} & \multicolumn{1}{l}{} & \multicolumn{1}{l}{} & \multicolumn{1}{l}{} & \multicolumn{1}{l}{}
\end{tabular}
\end{center}
\caption{
MSD challenge performance evaluation for lung tumor segmentation in terms of the Dice score and the normalized surface distance (NSD, higher is better). The metrics of our method are copied from the MSD leaderboard, and the metrics of other methods are copied from their respective publications and the leaderboard. *The results of U-Net$++$ are from the same task, but using a much larger training dataset.}
     \label{tab:task06}
\end{table}

\noindent\textbf{Comparison with nnU-Net:} nnU-Net has been validated as the state-of-the-art method for several medical image segmentation tasks~\cite{DBLP:journals/corr/abs-1809-10486,isensee2021nnu}.
It leverages 3 different types of 2D/3D U-Net architectures for prediction with ensembling.
It requires training 15 models (3 networks for 5-fold cross validation) in parallel to determine which models are used to create an ensemble and generate the final prediction.
Although we also conducted 5-fold cross validation using the searched architecture and training configuration, only one fixed neural architecture needs to be trained on each fold for a new task.
This is because once the searching on a single task is accomplished, it can be applied to other similar tasks directly.
Moreover, nnU-Net requires deep supervision during model training, which is unnecessary for our method.
Therefore, the marginal effort of applying our searched model and training configuration to a new task is considerable less than nnU-Net.
At the same time, our method achieved better performance and the  overall segmentation quality.

\subsection{Ablation Studies and Discussion}
\label{ablation_studies}

\begin{figure}[t]
    \centering
    \includegraphics[width=\linewidth]{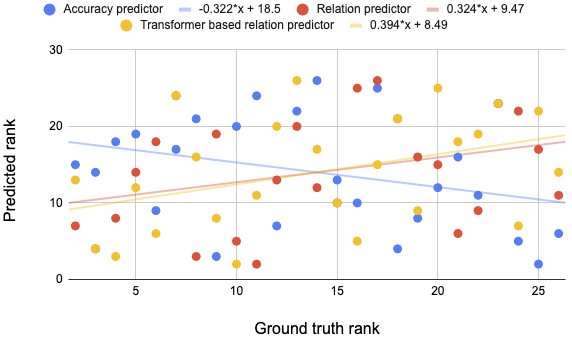}
    \caption{Comparison of different predictors for ranking prediction in validation set with 25 uniformly sampled configurations.}
    \label{fig:comp}
\end{figure}

We conduct ablation studies to motivate our predictor choice (see  Fig.~\ref{fig:comp}). For fair comparisons, we utilize three different types of predictors to rank data points in the validation set. The accuracy-based predictor generates accuracy for each validation data, and uses the accuracy to rank them.
The relation and transformer based relation predictor generates ranks after sorting.
The regular relation predictor is based on a multi-layer perceptron (MLP).
From Fig.~\ref{fig:comp}, we see the accuracy predictor cannot produce a positive correlation ($-0.322$) between ground truth ranks and predicted ranks (based on linear regression).
But both relation-based predictors can rank them in a positively correlated fashion.
Especially the relation-based predictor captures the top ranking points (left bottom corner of Fig.~\ref{fig:comp}).
And the transformer-based predictor has more than 20\% improvement over the MLP-based predictor in terms of the  correlation strength because of the advanced capacities of transformer modules.
Because the accuracy predictor has much fewer GT points used for training, the training set is easily overfitted, resulting in poor validation performance.
But the relation-based predictors leverage the pair-wise information (much more than individual accuracy) from the same training pool, and therefore are able to produce an improved ranking performance.





%
Based on the results of LiTS, apart from lesion segmentation, our method also works very well on organ segmentation.
However, search on organ segmentation may not be able to generate good architecture or training configurations for lesion segmentation (even in the same dataset).
This is because organ body segmentation is much simpler than lesion segmentation (especially for healthy cases).
And organ segmentation models typically converge much faster than lesion segmentation models.
Most of the segmentation networks with simple settings would achieve decent performance, even without data augmentation.
The liver segmentation model would converge within 2,000 iterations via a constant learning rate 0.001, and its validation accuracy achieves a Dice score above 0.950.
However, using the same dataset, the lesion segmentation model would take more iterations (around 10,000) to converge and data augmentation is necessary.
Because of the relative simplicity of liver segmentation, searched configuration did not fully exploit the potentials of models and training recipes.
Therefore, searching on liver segmentation would result in an AutoML predictor with low discriminatory capacity among candidates.
Hence, we suspect that challenging tasks (such as lesion segmentation) are more helpful for AutoML search in order to transfer the searched results to other applications.

%
\section{Conclusion}
\label{conclusion}
In this paper, we proposed a novel AutoML method to optimize deep learning configurations for lesion segmentation in 3D medical images leveraging the advanced capacity of transformer modules. The proposed method predicts the relation between different training configurations and neural networks.
We introduced a new search space for the neural architecture through modifications of SpineNet~\cite{DBLP:conf/cvpr/DuLJGTCLS20}.
Meanwhile, we created a predictor-based AutoML algorithm, which is computationally efficient and effective.
It can cover ``almost'' all components in conventional deep learning frameworks.
Our experiments showed that, compared to other existing methods in the literature, our method can achieve the most advanced performance in large-scale lesion segmentation datasets.
Furthermore, the proposed methods have been proven to be effectively transferable to different datasets.
Future work could investigate how to further reduce searching and training time given certain computational budgets.

{\small
    \bibliographystyle{unsrt}
    \bibliography{main_final}
}
\end{document}